\documentclass[aps,pre,twocolumn,amsmath,amssymb,superscriptaddress,reprint,longbibliography]{revtex4-1}
\usepackage{amsfonts,amsmath,amssymb,bm}
\usepackage{graphicx,graphics,float}
\usepackage{bm}
\usepackage{amssymb}
\usepackage{colordvi}
\usepackage{graphicx}
\usepackage{color}
\usepackage[colorlinks=true,linkcolor=blue,citecolor=blue,urlcolor=blue]{hyperref}%
\usepackage{hyperref}
\usepackage{comment}
\usepackage{harpoon}
\usepackage{braket}
\begin{document}

\title{Coherence-enhanced thermodynamic performance in a  periodically-driven inelastic heat engine}

\author{Jincheng Lu}
\address{Jiangsu Key Laboratory of Micro and Nano Heat Fluid Flow Technology and Energy Application, School of Physical Science and Technology, Suzhou University of Science and Technology, Suzhou, 215009, China}

\author{Zi Wang}
\address{Center for Phononics and Thermal Energy Science, China-EU Joint Center for Nanophononics, Shanghai Key Laboratory of Special Artificial Microstructure Materials and Technology, School of Physics Science and Engineering, Tongji University, Shanghai 200092 China}

\author{Jie Ren}
\address{Center for Phononics and Thermal Energy Science, China-EU Joint Center for Nanophononics, Shanghai Key Laboratory of Special Artificial Microstructure Materials and Technology, School of Physics Science and Engineering, Tongji University, Shanghai 200092 China}

\author{Chen Wang}\email{wangchen@zjnu.cn}
\address{Department of Physics, Zhejiang Normal University, Jinhua, Zhejiang 321004, China}

\author{Jian-Hua Jiang}\email{joejhjiang@hotmail.com}
\affiliation{Suzhou Institute of Advanced Research, University of Science and Technology of China, Suzhou, 215123, China}

\date{\today}
\begin{abstract}
Quantum thermodynamics with microscopic inelastic scattering processes has been intensively investigated in recent years.
Here, we apply quantum master equation combined with full counting statistics approach to investigate the role of quantum coherence on the periodically-driven inelastic heat engine. 
 We demonstrate that the inelastic quantum heat engine exhibits dramatic advantage of thermodynamic performance compared to their elastic counterpart.
 Moreover, it is found that  inelastic currents, {output work}, and the efficiency can be enhanced by quantum coherence. 
 In particular, the geometric effect proves crucial in achieving maximal values of generated output work and energy conversion efficiency. 
 The Berry curvature boosted by quantum coherence unveils the underlying mechanism of  periodically-driven inelastic heat engine. 
Our findings may provide some insights for further understanding and optimizing periodically-driven heat engines
via quantum coherence resource and inelastic scattering processes.

\end{abstract}

\maketitle

\section{Introduction}

Quantum thermodynamics, an exquisite combination of  thermodynamics and quantum mechanics, addresses
heat-to-work conversion and entropy production in quantum thermal machines at the microscopic level, ranging from the heat engines, refrigerators, heat pumps, and even multitask machines~\cite{kosloff2013entropy,anders2016cp,BenentiPR,Arrachea23}. The practical quantum thermodynamics mainly considers nonequilibrium thermodynamic processes, which are typically realized by
(i) time-dependent modulations~\cite{SeifertPR,lacerda2023prb}; 
(ii) multiple reservoirs with thermodynamic bias~\cite{jauho2008book,Nanotechnology,JiangCRP};
(iii) quantum measurements~\cite{BuffoniPRL,HasegawaPRL20}; (iv) quantum information, e.g., quantum correlation~\cite{demon1,lutz2019nc,SahaPRL}.
The periodically driven quantum heat engines have attracted increasing attention, which can overcome thermodynamic biases to
sustain the heat transfer from the cold (low voltage) drain to the hot (high voltage) source, thereby enabling the thermodynamic operations~\cite{ArracheaPRB20,JunjiePRL,CangemiPRB,MyPRBGTUR,ErdmanPRR23,cavaliere23,WangPRR23,HinoPRR21}. In particular, the geometric effects~\cite{bohm03,ziwang2022fop}, e.g., Berry phase and quantum  metric, should be properly adopted.

Quantum coherence is one kind of indispensable ingredients for quantum mechanics and also a fundamental quantum resource in quantum thermodynamics, which distinguish from classical counterparts~\cite{plenio2014prl,StreltsovRMP}.
With its unique features, quantum coherence finds fertile applications in quantum thermal machines~\cite{WackerPRB11,UzdinPRX15,SuPRE16,BrandnerPRL17,Niedenzu,PRB18coherence,CamatiPRA19,FrancicaPRL,HammamNJP21,SegalPRE21}. 
Notably, heat engine, heat pump, and multitask thermal machine can be driven by pure quantum coherence~\cite{kawai2022pra,KwonPRA23,Manzano23}. Quantum coherence can also enhance the efficiency and constancy of the quantum thermal machines~\cite{Brandner15NJP}. Moreover, quantum coherence enables us to explore the quantum contribution to the nonequilibrium entropy production 
and  information processes, e.g., nonequilibrium Landauer principle~\cite{landi2023rmp,SaitoPRL22}.

Recently, there has been a growing recognition of the significance of inelastic scattering processes in nonequilibrium transport and thermodynamics~\cite{Jiang2012,Jiang2013,Yamamoto,MyReview,NianPRB23},  which are implemented in three-terminal setups, in contrast to the elastic scattering processes sufficiently realized via two terminals.
The generic inelastic processes enable one to investigate nonlinearly-coupled electronic and bosonic currents.
Interestingly, it is found that the bounds of Onsager coefficient with inelastic processes are dramatically relaxed to promote the thermodynamic performance~\cite{MyPRBtransistor}.
Thus, quantum thermal machines (e.g., thermal transistor and refrigerator)  exhibit thermodynamic advantage by comparing with counterparts under elastic processes. 
Meanwhile,  such  microscopic inelastic processes yield other unconventional transport and thermodynamic phenomena, e.g., cooling by heating~\cite{Cooling2}, the separation of charge and heat currents~\cite{OraPRB2010,MazzaPRB}, linear transistor effects~\cite{Jiangtransistors,MyPRBdiode},
and cooperative heat engines~\cite{JiangJAP}.
Though extensive studies have been conducted to excavate the steady-state thermodynamics of inelastic quantum thermal machines~\cite{BijayJiang}, the influence of interplay between quantum coherence and geometric effects on thermodynamic performance of periodically-driven quantum heat engines are far from clear.

In our work, we have conducted a comprehensive study to investigate the impact of quantum coherence on  thermodynamic performance of a periodically driven  heat engine modulated in adiabatic regimes, by including the Redfield equation in absence of secular approximation.
First, the performance of a three-terminal inelastic quantum heat engine is compared with a two-terminal elastic counterpart to demonstrate the advantages of inelastic scattering processes. 
Then, the geometric and dynamic current components dramatically affected by quantum coherence are obtained,  and the contribution of the geometric component to the thermodynamic performance of three-terminal heat engine is rigorously analyzed.

The main points of this work are demonstrated as: 
(i) The Redfield equation encompasses a unified description of nonlinear electronic and phononic transport in a three-terminal periodically-driven setup, with a particular focus on the interplay between inelastic electron-phonon scattering and quantum coherence on nonequilibrium currents.
Thus the method goes beyond the traditional transport approaches, e.g., Fermi-golden rule~\cite{MillerPR60,Jiang2012}, the Lindblad equation~\cite{lindblad,TupkaryPRA}, and Redfield equation with secular approximation~\cite{REDFIELD19651}.
(ii)  The three-terminal driven inelastic quantum heat engine  exhibits significant advantage of thermodynamic performance in a wide parameter regime, compared to two-terminal elastic counterpart~\cite{JuergensPRB13,JordanPRR22}.
This clearly demonstrates the importance role of the inelastic scattering processes on driven quantum thermodynamics. 
Additionally, the combination of heat and particle transports in inelastic transport facilitates concurrent optimization of thermal conductance and electrical conductance for thermodynamic devices.    
(iii) The thermodynamic performance of the driven inelastic heat engine can be dramatically enhanced via quantum coherence. The nonequilibrium currents are decomposed as  dynamic and geometric components. And the geometric component is found to be crucial in achieving maximal values of generated output work and thermodynamic efficiency.

This study is structured as follows: In Sec.~\ref{sec:model}, we describe the setup of the inelastic heat engines and derive the dynamic, geometric currents using the quantum master equation with a full counting statistics approach. The work and efficiency also have been defined.  
{In Sec.~\ref{sec:results}, we focus on analyzing the energy efficiency and output work of the inelastic heat engines. Additionally, we numerically compare the results with both the elastic engine and the incoherent engine.} We summarize our findings in Sec.~\ref{sec:con}. For simplicity, we set $\hbar=k_B=e\equiv 1$ throughout this paper.

\begin{figure}[htb]
\begin{center}
\centering\includegraphics[width=8.5cm]{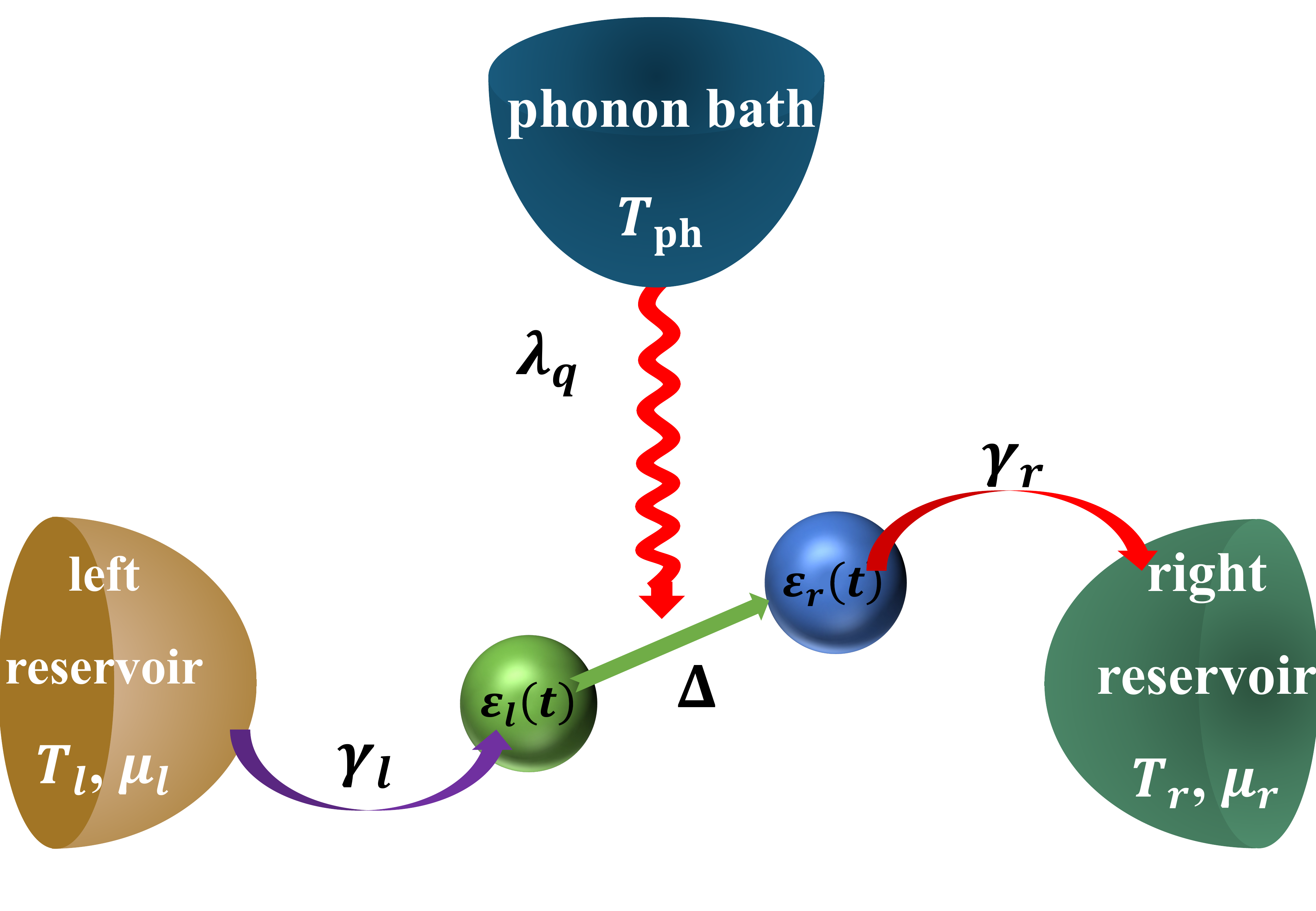}
\caption{Illustration of the three-terminal inelastic heat engine: An electron initially departs from the left reservoir and enters the left QD, characterized by an energy level $\varepsilon_l $. Subsequently, the electron undergoes a transition to the right QD, which possesses a different energy level denoted as $\varepsilon_r $. This transition is facilitated by interacting with one phonon from the phonon bath, maintained at a temperature of $T_{\rm ph}$. Within this setup, two electric reservoirs are involved, each characterized by distinct temperatures and chemical potentials. The left ($l$) and right ($r$) electric reservoirs have temperatures denoted as $T_{l(r)}$, while their respective chemical potentials are represented as $\mu_{l(r)}$. $\Delta$ represents the tunneling strength between two QDs, $\gamma_i$ characterizes the coupling between the electronic reservoirs and the corresponding QD, and $\lambda_q$ represents the electron-phonon interaction strength.}\label{fig:3Tsystem}
\end{center}
\end{figure}

\section{Model and methods}~\label{sec:model}

\subsection{Inelastic heat engine}~\label{sec:DQDsystem}

We consider an inelastic heat engine, which is composed of a double quantum dot (QD) system inelastically coupled to a phonon bath, 
and each dot individually exchanges energy with an electronic reservoir (e.g., metal lead, denoted as $l$ and $r$), as shown in Fig.~\ref{fig:3Tsystem}. 
The Hamiltonian of this inelastic heat engine reads
$\hat H = \hat H_{\rm DQD} + \hat H_{\rm e-ph} + \hat H_{\rm lead} + \hat H_{\rm tun} + \hat H_{\rm ph}$~\cite{MyReview}.
Specifically, the double QDs is described as
 \begin{equation}
 \hat H_{\rm DQD} = \sum_{i=l,r} \varepsilon_i \hat d_i^\dagger \hat d_i +  {\Delta}(\hat d_l^\dagger \hat d_r + {\rm H.c.}), 
 \end{equation}
where $\hat{d}_i^\dagger$ ($\hat{d}_i$) is the creation (annihilation) operator of one electron in the $i$-th QD, $\varepsilon_i$ represents the QD energy, and $\Delta$ shows tunneling between the two QDs.
The phonon reservoir denotes
$\hat H_{\rm ph} = \sum_q\omega_{q}\hat a^\dagger_q \hat a_q$, where  $\hat{a}^\dagger_q$ ($\hat{a}_q$) creates (annihilates) one phonon with the frequency $\omega_{q}$.
The inelastic electron-phonon interaction  is described as
\begin{eqnarray}
\hat H_{\rm e-ph} = \sum_{q}\lambda_q \hat d_l^\dagger \hat d_r (\hat a_q + \hat a^\dagger_q) + {\rm H.c.},
\end{eqnarray}
where  $\lambda_q$ is the strength of the electron-phonon coupling strength. 
The electronic leads are expressed as
$\hat H_{\rm lead} = \sum_{j=L,R}\sum_{k} \varepsilon_{jk} N_{jk}$, 
with the electron number $N_{jk} = \hat d_{jk}^\dagger \hat d_{jk}$ in the $j$-th lead at the momentum $k$.
The electron tunneling between the dots and the electronic reservoirs are given by
$ \hat H_{\rm tun} = \sum_{j=L,R;k} \gamma_{jk} \hat d_j^\dagger \hat d_{jk} +  {\rm H.c.}$,
where $\gamma_{jk}$  is the corresponding coupling strength.

To analyze the heat engine in  eigenspace of the double QDs,
we begin to diagonalize $\hat H_{\rm DQD}$ as
\begin{equation}
\hat H_{\rm DQD} =  E_D\hat D^\dagger\hat D + E_d\hat d^\dagger\hat d, 
\end{equation} 
where the eigenenergies denote
$E_D = \frac{\varepsilon_r +\varepsilon_l }{2}+\sqrt{\frac{(\varepsilon_r -\varepsilon_l )^2}{4}+\Delta^2}$ and $E_d = \frac{\varepsilon_r +\varepsilon_l }{2}-\sqrt{\frac{(\varepsilon_r -\varepsilon_l )^2}{4}+\Delta^2}$,
and the new sets of Fermion operators are specified as
$\hat D = \sin\theta \hat d_l + \cos\theta \hat d_r$ and $\hat d = \cos\theta \hat d_l - \sin\theta \hat d_r$~\cite{MyPRBdiode,JordanPRR22}, with $\theta\equiv \arctan\left(\frac{2\Delta}{\varepsilon_r -\varepsilon_l }\right)/2$.
Consequently, the electron-phonon and dot-reservoir tunneling terms are reexpressed as
$\hat H_{\rm e-ph} =   \sum_q\lambda_q [\sin(2\theta)(\hat D^\dagger\hat D-\hat d^\dagger\hat d) + \cos(2\theta)(d^\dagger D+D^\dagger d)](\hat a^\dagger_q + \hat a_q)$
and
$ \hat H_{\rm tun} = \sum_k [\gamma_{Lk} (\sin\theta\hat D^\dagger + \cos\theta\hat d^\dagger) \hat d_{Lk} 
+  \gamma_{Rk} (\cos\theta\hat D^\dagger - \sin\theta\hat d^\dagger) \hat d_{Rk}] +  {\rm H.c.}$. 
From the term $\hat{H}_{\rm e-ph}$ it is known that in the eigenbasis of $\hat{H}_{\rm DQD}$ there both exist dephasing and damping processes, which may generate the steady-state coherence~\cite{purkayastha20npj}.


\subsection{Geometric-phase-induced currents}


Full counting statistics nowdays is widely accepted as a powerful utility to characterize complete information of current fluctuations~\cite{fluctuationRMP}. Based on two-time measurement protocol~\cite{CampisiRMP},
we apply the full counting statistics to obtain the particle and energy flows out of electronic reservoirs and heat current out of phonon reservoir by including 
$\mathbf{\Lambda}=\{\lambda_p$, $\lambda_E$, $\lambda_{\rm ph}\}$, respectively(see the introduction of full counting statistics in Appendix \ref{sec:FCS}).  
Consequently, the counting-field-dependent total Hamiltonian is described as    
\begin{equation}
H_{-\mathbf{\Lambda}/2} 
= H_{\rm DQD} + H_{\rm ph} + H_{\rm lead} + V_{-\mathbf{\Lambda}/2}, 
\end{equation}
with $V_{-\mathbf{\Lambda}/2}$ specified as         
\begin{equation}
\begin{aligned}
V_{-\mathbf{\Lambda}/2} &= \sum_q\lambda_q [\sin(2\theta)(\hat D^\dagger\hat D-\hat d^\dagger\hat d) \\
&+ \cos(2\theta)(d^\dagger D+D^\dagger d)](e^{i\frac{\lambda_{\rm ph}}{2}\omega_q}{\hat a}_q +  {\rm H.c.}) \\
&+ \sum_k ([\gamma_{Lk} (\sin\theta\hat D^\dagger + \cos\theta\hat d^\dagger) \hat d_{Lk}  \\
&+  \gamma_{Rk}e^{-i\frac{\lambda_p}{2}-i\frac{\lambda_E}{2}\varepsilon_{Rk}} (\cos\theta\hat D^\dagger - \sin\theta\hat d^\dagger) \hat d_{Rk}] +  {\rm H.c.}).  
\end{aligned}
\end{equation}
We assume the electron-phonon coupling and dot-reservoir tunnelings are weak. Based on the Born-Markov approximation,
we perturb $V_{-\mathbf{\Lambda}/2}$ to obtain the quantum master equation as~\cite{breuer02}
\begin{equation}
\begin{aligned}
&\frac{\partial}{\partial t} \rho_S(\mathbf{\Lambda},t)
= i[\rho_S(\mathbf{\Lambda},t),H_{\rm DQD}] \\
&- \int_0^\infty d\tau {\rm  Tr}_B\{[
[V_{-\mathbf{\Lambda}/2},[V_{-\mathbf{\Lambda}/2}(-\tau),
\rho_S(\mathbf{\Lambda},t){\otimes}\rho_B]_{\mathbf{\Lambda}}
]_{\mathbf{\Lambda}}\}, 
\label{eq:rhoSV}
\end{aligned}
\end{equation}
where $\rho_S(\mathbf{\Lambda},t)$ denotes the reduced density operator of {central double QDs system} with counting parameters, i.e.
$\rho_S(\mathbf{\Lambda},t)
=\textrm{Tr}_B\{\rho^T_{\mathbf{\Lambda}}(t)\}$,
with $\rho^T_{\mathbf{\Lambda}}(t)$ (see Eq.~\eqref{eq:A4} in Appendix \ref{sec:FCS}) the full density operator of the whole inelastic heat engine,
the commutating relation denotes
$[\hat{A}_{\mathbf{\Lambda}},\hat{B}_{\mathbf{\Lambda}}]_{\mathbf{\Lambda}}
=\hat{A}_{\mathbf{\Lambda}}\hat{B}_{\mathbf{\Lambda}}
-\hat{B}_{\mathbf{\Lambda}}\hat{A}_{-\mathbf{\Lambda}}$,
and the equilibrium distribution of reservoirs is specified as
$\rho_B=\rho_l{\otimes}\rho_r{\otimes}\rho_{\rm ph}$,
with 
$\rho_i=\exp\left[-\beta_{i}(\hat H_{i} - \mu_i\hat N_i)\right] / Z_i~(i=l, r)$,
$\rho_{\rm ph}=\exp\left[-\beta_{\rm ph}\hat H_{\rm ph} \right] / Z_{\rm ph}$,
the partition functions 
 $Z_i = \text{Tr}\left\{\exp\left[-\beta_{i}(\hat H_{i} - \mu_i\hat N_i)\right]\right\}$ 
 and
$Z_{\rm ph} = \text{Tr}\left\{\exp\left[-\beta_{\rm ph}\hat H_{\rm ph} \right]\right\}$,
and the inverse temperatures
$\beta_i=1/k_BT_i$ and $\beta_{\rm ph}=1/k_BT_{\rm ph}$.
If we reorganize $\rho_S(\mathbf{\Lambda},t)$ in the vector form
$| {\mathbf P}(\mathbf{\Lambda},t) \rangle$ = $[\langle 0 |\rho_S (\mathbf{\Lambda},t) | 0 \rangle$; $\langle D |\rho_S (\mathbf{\Lambda},t) | D \rangle$; $\langle d |\rho_S (\mathbf{\Lambda},t) | d \rangle$; $\langle D |\rho_S (\mathbf{\Lambda},t) | d \rangle $; $\langle d |\rho_S (\mathbf{\Lambda},t) | D \rangle ]$,
the quantum master equation is reexpressed as
\begin{equation}
\begin{aligned}
\dfrac{d | {\mathbf P}({\mathbf{\Lambda},t}) \rangle} {dt}= {\mathbf H}(\mathbf{\Lambda},t)| {\mathbf P}(\mathbf{\Lambda},t)\rangle, 
\end{aligned}
\end{equation}
where  $ {\mathbf H}(\mathbf{\Lambda},t)$ is the evolution matrix with its elements shown in Appendix~\ref{countright}. 
We note that the inclusion of the off-diagonal elements of the density matrix, i.e., $\langle D |\rho_S (\mathbf{\Lambda},t) | d \rangle $ and $\langle d |\rho_S (\mathbf{\Lambda},t) | D \rangle $,
are the signature of quantum coherence.
Conversely, if we neglect the quantum coherence effect, we disregard the off-diagonal elements.


For the heat engine modulated by parameters such as $\Gamma_i(t)$, $\mu_i(t)$, $T_i(t)$, and $\varepsilon_i(t)$ ( $i=l,r$), in the long time evolution the cumulant generating function based on the large deviation principle and adiabatic perturbation theory can be divided into two components~\cite{RenPRL10}
\begin{eqnarray}
\mathcal{G}_{\rm tot}(\mathbf{\Lambda})=\mathcal{G}_{\rm dyn}(\mathbf{\Lambda})+\mathcal{G}_{\rm geo}(\mathbf{\Lambda}),
\end{eqnarray}
where the dynamical phase denotes $\mathcal{G}_{\rm dyn}(\mathbf{\Lambda})=\frac{1}{\mathcal T}\int_0^{{\mathcal T}}dt E_g(\mathbf{\Lambda},t)$, with $E_g(\mathbf{\Lambda},t)$ the eigenvalue of 
$ {\mathbf H}(\mathbf{\Lambda},t)$ owning the maximum real part~\cite{Sinitsyn07,RenPRL10}, and the geometric phase is specified as
${\mathcal G}_{\rm geo}(\mathbf{\Lambda}) =- \frac{1}{\mathcal T}\int_0^{{\mathcal T}}dt \langle \varphi(\mathbf{\Lambda},t)|\partial_t|\psi(\mathbf{\Lambda},t)\rangle$,
with $\langle \varphi(\mathbf{\Lambda},t)|$
and
$|\psi(\mathbf{\Lambda},t)\rangle$ left and right eigenvectors of ${\mathbf H}(\mathbf{\Lambda},t)$.
 The dynamical component, ${\mathcal G}_{\rm dyn}$, characterizes the temporal average and delineates the dynamic aspects of particle and heat transfer. While the geometric contribution, ${\mathcal G}_{\rm geo}$, arises from adiabatic cyclic evolution and necessitates a minimum of two parameter modulations to manifest its effects~\cite{Sinitsyn07,RenPRL10,NiePRA20}.

Consequently, the particle current flowing from the right reservoir into the system is given by~\cite{BijayJiang}:
\begin{equation}
\langle N_l  \rangle = \frac{\partial{\mathcal G_{\rm tot}(\mathbf{\Lambda})}}{\partial (i\lambda_p)}\Big{|}_{\mathbf{\Lambda}=0},
\end{equation}
and the particle flow from the left reservoir 
can be straightforwardly obtained as
$\langle N_l \rangle =- \langle N_r \rangle$
via the law of particle conservation.
Meanwhile the energy flow is expressed as
\begin{equation}
\langle E_r \rangle  = \frac{\partial{\mathcal G_{\rm tot}(\mathbf{\Lambda})}}{\partial (i\lambda_E)}\Big{|}_{\mathbf{\Lambda}=0},
\end{equation}
and electronic heat flow extracted from the right reservoir is defined as
$\langle Q_r \rangle = \langle E_r \rangle  - \mu_r \langle N_r \rangle$~\cite{YamamotoPRE15}.
The phononic heat current is given by
\begin{equation}
\langle Q_{\rm ph}\rangle  = \frac{\partial{\mathcal G_{\rm tot}(\mathbf{\Lambda})}}{\partial (i\lambda_{\rm ph})}\Big{|}_{\mathbf{\Lambda}=0}.
\end{equation}
In analogy, the particle current $\langle N_l\rangle$ and energy current $\langle E_l\rangle$ flowing from the left electronic reservoir into the central system. In this study, we constrain all parameters of the driving protocol to an adiabatic driving regime: the driving period is chosen as $\mathcal T = 10^{-12}$ s, corresponding to $\hbar\Omega \approx 4\times 10^{-2}$ meV. It's evident that the rate between system and reservoir $\Gamma_{i} = 4$ meV  ($i=l,r,{\rm ph}$), which is much greater than $\hbar\Omega$, and the adiabatic approximation remains valid~\cite{MyPRBGTUR}.  In this work, we demonstrate the realization of the thermoelectric engine in the double quantum dot system by choosing the left and right quantum dot energies, i.e., $\varepsilon_l $ and $\varepsilon_r $, as the modulating parameters~\cite{JuergensPRB13,HinoPRR21,GuoPRL23}.  The quantum dots are driven adiabatically following the protocol: $\varepsilon_l = E_{A,l}+E_{B,l}\sin(\Omega t)$, $\varepsilon_r = E_{A,r}+ E_{B,r}\sin(\Omega t + \phi)]$. 
{ The impact of geometric properties can be realized by tuning the phase $\phi$. 
Nonzero modulation phase $\phi$ is required to pump heat from one reservoir to another reservoir.
Distinct values of $\phi$ correspond to different driving protocols. 
When $\phi=\pi/2$, the modulation induced geometric pump is optimized. In contrast, when $\phi=0$, the geometric contribution diminishes, leaving only the dynamical counterpart.
This stems from the disappearance of
the enclosed area in the parameter space, e.g., $\varepsilon_{l}$ and $\varepsilon_{r}$.
}

\subsection{Definitions of work and efficiency}

We operate the three-terminal inelastic heat engine by harvesting energy from the hot phonon reservoir with a fixed temperature of $k_BT_l=k_BT_r=10\, {\rm meV}$ and $k_BT_{\rm ph}=12\, {\rm meV}$ and converting it into useful output work. The electrochemical potential bias is defined as $\Delta\mu=\mu_l-\mu_r$, with the average chemical potential $\mu \equiv (\mu_l+\mu_r)/2$. 
The particle and energy conservation laws imply that~\cite{JunjiePRL, MyPRBGTUR}, 
\begin{equation}
 \langle W_I \rangle=-(\braket{ E_l} + \langle E_r \rangle + \langle Q_{\rm ph} \rangle) .
\label{eq:conversation}
\end{equation}
Here, $\langle W_I \rangle$ represents the input work per modulating period $\mathcal T$,
which becomes vanishing once the driving is removed.
And the useful output work of the heat engine is described as
\begin{equation}
\braket{W_{\rm out}} = (\mu_l-\mu_r) \braket{N_r}.
~\label{eq:power}
\end{equation}
The entropy production of the whole system is given by $\langle S \rangle=-\sum_{v=l,r,\rm ph} \langle Q_v\rangle/T_v$~\cite{MyPRBMultitask}. Moreover, the entropy production takes on a specific form~\cite{MyPRBMultitask}
\begin{equation}
T_l\langle S \rangle = \left(1- {T_l/T_{\rm ph}}\right)\langle Q_{\rm ph}\rangle-\langle W_{\rm out}\rangle+\langle W_I\rangle.
\label{eq:entropy}
\end{equation}
From the above equation, we can find that the thermodynamic device can still function as a quantum heat engine and generate useful work, even without a temperature difference.  We note that when the electric power $\braket{W_{\rm out}}>0$, the thermal machine operates as a heat engine. 
(i) If the input energy is negative, i.e., $\langle W_I\rangle <0$, the efficiency of the heat engine is given by
\begin{equation}~\label{meta}
\langle \eta \rangle= \frac{\langle W_{\rm out}\rangle}{(1 - {T_l/T_{\rm ph}})\langle Q_{\rm ph}\rangle}.
\end{equation}
Such definition of the efficiency is consistent with the energy efficiency of steady-state thermoelectric transport,
e.g., at the Carnot limit ${\langle W_{\rm out}\rangle}/\langle Q_{\rm ph}\rangle=1-{T_l/T_{\rm ph}}$,
the efficiency at Eq.~(\ref{meta}) becomes the unit. 
(ii) When the input energy is nonnegative, i.e. $\langle W_I\rangle {\ge}0$, the efficiency of the heat engine becomes~\cite{MyPRBGTUR}
\begin{equation}
\langle \eta \rangle= \frac{\langle W_{\rm out}\rangle}{(1 - {T_l/T_{\rm ph}})\langle Q_{\rm ph}\rangle + \langle W_I\rangle}. \label{eq:meta2}
\end{equation}
According to the thermodynamic second law, the thermoelectric engine efficiency is always upper bounded by  $\langle\eta\rangle \le 1$~\cite{JiangPRE,JunjiePRL}.

In contrast for the elastic heat engine, it is known that ${\langle}Q_{\rm ph}{\rangle}=0$ in absence of the inelastic electron-phonon scattering. Thus, the average input work is reduced to
$\langle W_I \rangle=-(\braket{ E_l} + \langle E_r \rangle)$. 
Meanwhile, the entropy production of Equation~\eqref{eq:entropy}  is simplified to
$T_l\langle S \rangle = -\langle W_{\rm out}\rangle+\langle W_I\rangle$,
and the thermodynamic efficiency denotes
\begin{eqnarray}
   {\braket \eta}=\theta(\langle W_I\rangle)\langle W_{\rm out}\rangle/\langle W_I\rangle,
\end{eqnarray}
with $\theta(x)=1$ for $x>0$ and $\theta(x)=0$ for $x{\leq}0$.
 It should note that the input energy of an elastic quantum heat engine originates entirely from the driving energy $W_I=-(\braket{E_l}|{\rm geo} + \langle E_r \rangle |{\rm geo})$, and $\braket{E_l}|{\rm dyn} + \langle E_r \rangle |{\rm dyn}\equiv0$ due to the second law of thermodynamics~\cite{JiangCRP}.

\section{Results and discussions}\label{sec:results}
In this section, we will present results derived from periodically-driven double quantum dot setups. Our investigation will encompass two distinct cases: (i) a comparison of thermodynamic performance between elastic and inelastic heat engines with quantum coherence, and (ii) an exploration of the impact of quantum coherence on the inelastic engine. 

\subsection{Elastic vs inelastic periodically-driven heat engines}

\begin{figure}[htb]
\begin{center}
\centering\includegraphics[width=8.5cm]{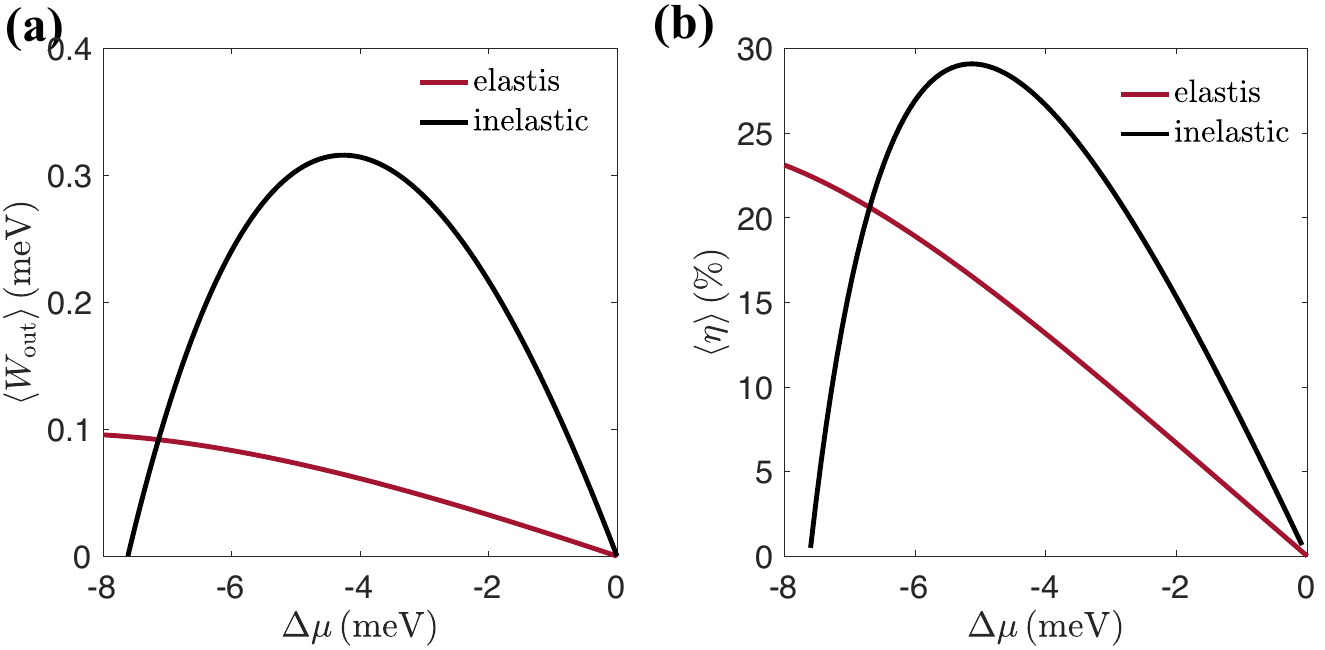}
\caption{Comparison of performance between elastic and inelastic quantum heat engines. (a) The average output electric work $\braket{W_{\rm out}}$, (b) average  efficiency $\braket{\eta}$ as a function of $\Delta\mu$ for two-terminal (elastic) and three-terminal (inelastic) heat engines. The energy lelvel modulations are exemplified as $\varepsilon_l= [-1+5\sin(\Omega t)]\, {\rm meV}$, $\varepsilon_r =[1 + 5\sin(\Omega t+\pi/2)]\, {\rm meV}$, $\Omega=2\pi/{\mathcal T}$ and ${\mathcal T}=10^{-12} \,{\rm s}$. 
The coupling between quantum dots and phonon reservoir for the inelastic and elastic cases are set as
$\Gamma_{\rm ph}=4 \, {\rm meV}$
and
$\Gamma_{\rm ph}=0$, respectively.
The other parameters are given by $\mu=0$, $\Gamma_l=\Gamma_r=4 \, {\rm meV}$, $\Delta=8 \, {\rm meV}$, $k_BT_l=k_BT_r=10\, {\rm meV}$, and $k_BT_{\rm ph}=12 \, \rm meV$. }\label{fig:elastic}
\end{center}
\end{figure}

For the periodically-driven elastic heat engines, a straightforward example of such devices is a two-terminal double quantum dots device, where the energy exchange between quantum dots and the phonon reservoirs is isolated,
i.e. $\Gamma_{\rm ph}=0$. 
We depict the average work and efficiency of both elastic (the red solid curve) and inelastic (the black solid curve) heat engines as a function of the potential difference $\Delta\mu$ in Fig.\ref{fig:elastic}, utilizing Eq.\eqref{eq:power} and Eqs.~\eqref{meta}-\eqref{eq:meta2} to compare their performances. 
Here quantum coherence is considered for both heat engines. 
It is found that even without a temperature difference ($T_l=T_r$), the devices can still work as a quantum heat engine and generate useful work, i.e., $\braket{W_{\rm out}}>0$.

{Fig.~\ref{fig:elastic} clearly shows that in a broad voltage bias range (as indicated by the red and black curves for $-8 \, \text{meV}<\Delta\mu<0$), the inelastic quantum heat engine demonstrates substantially higher output work and efficiency compared to its elastic counterpart. This superior performance is attributed to the benefits of inelastic scattering processes, such as inelastic thermoelectricity, which surpasses conventional elastic transport, as proposed by Mahan and Sofo for conventional thermoelectricity~\cite{Mahan}. Consequently, we assert that inelastic heat engines incorporating quantum coherence warrant thorough analysis.}

\begin{figure}[htb]
\begin{center}
\centering\includegraphics[width=8.5cm]{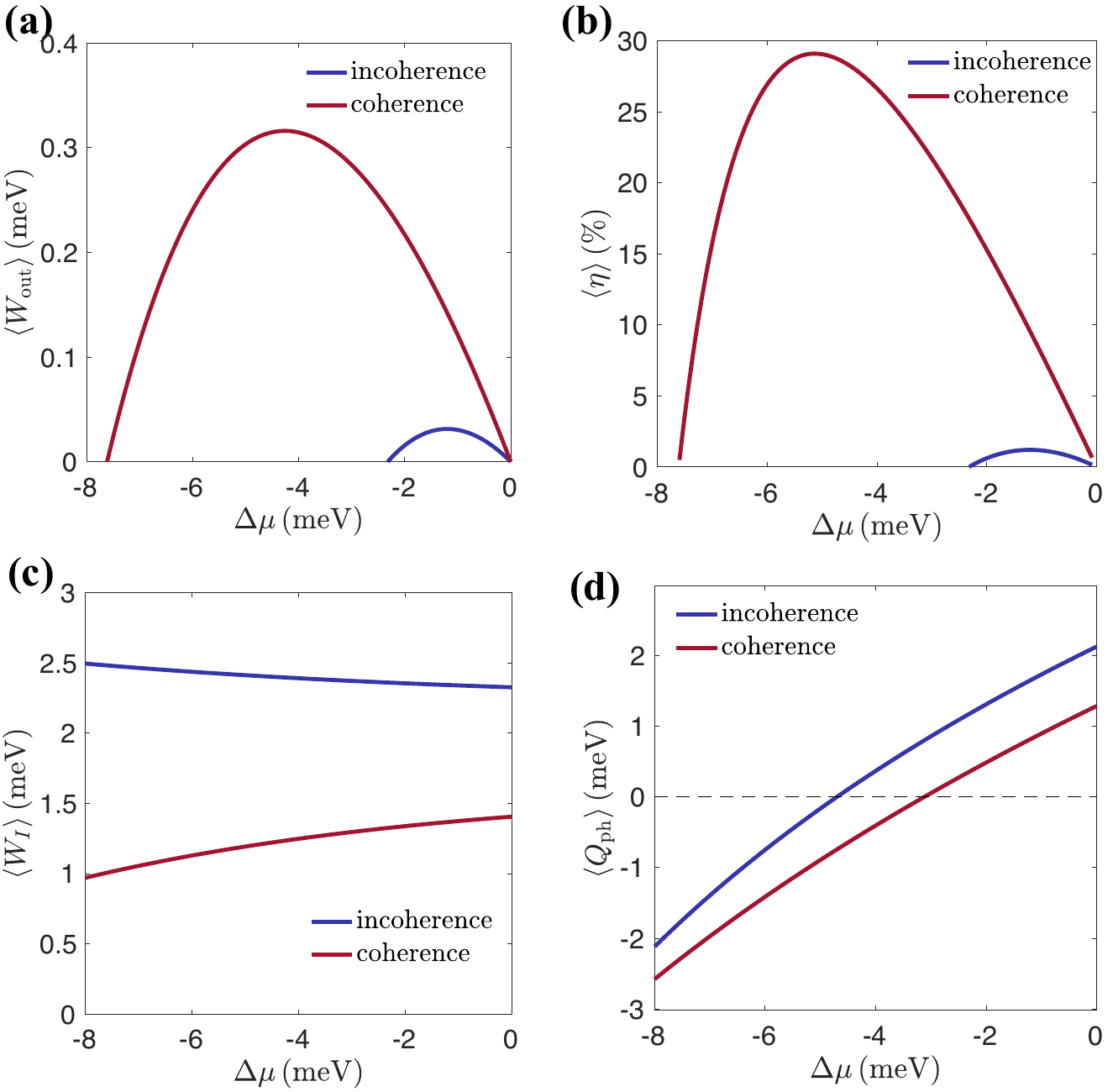}
\caption{(a) The average output electric work $\braket{W_{\rm out}}$, (b) average  efficiency $\braket{\eta}$, (c) the 
input work done by driving $\langle W_I \rangle$, and (d) average phononic current $\braket{Q_{\rm ph}}$  as a function of $\Delta\mu$ for the coherence and incoherence cases. The energy level modulations are exemplified as $\varepsilon_l= [-1+5\sin(\Omega t)]\, {\rm meV}$, $\varepsilon_r =[1 + 5\sin(\Omega t+\pi/2)]\, {\rm meV}$, $\Omega=2\pi/{\mathcal T}$ and ${\mathcal T}=10^{-12} \,{\rm s}$. The other parameters are given by $\mu=0$, $\Gamma_l=\Gamma_r=\Gamma_{\rm ph}=4 \, {\rm meV}$, $\Delta=8 \, {\rm meV}$, $k_BT_l=k_BT_r=10\, {\rm meV}$, and $k_BT_{\rm ph}=12 \, \rm meV$. }\label{fig:WetaQph}
\end{center}
\end{figure}

\subsection{Thermodynamic performance of the inelastic heat engine}~\label{sec:performance}
In Fig.~\ref{fig:WetaQph}, we commence the analysis of the thermodynamic performance of the inelastic engine employing a three-terminal double QDs system. 
It need note that when we naturally include quantum coherence, we mark the calculated quantities (e.g., the output work) with ``coherence".
In contrast if we discard quantum coherence, these quantities are marked with ``incoherence".
Initially, we examine the output work and efficiency of the engine, depicted in Fig.~\ref{fig:WetaQph}(a) and \ref{fig:WetaQph}(b), respectively.
We observe that the nonzero quantum coherence (off-diagonal elements of system density matrix) yields a significant improvement in the optimal work and efficiency, compared with the counterparts in absence of quantum coherence. 
More specifically, the maximum output work increases to $0.3\, \rm meV$ as the coherence effect is taken into account, whereas it becomes $0.05\, \rm meV$ by artificially ignoring quantum coherence, {see Fig.~\ref{fig:WetaQph}(a)}. 
In analogy, the maximum efficiency considering the coherent transport effect lead to a maximum efficiency of $30\%$ (in percentage units), whereas it becomes only $2\%$
during the incoherent transport processes, {as shown in Fig.~\ref{fig:WetaQph}(b)}. (More details of the optimization and comparison can be found in Appendix \ref{sec:appC}.) 
It's worth noting that the temperatures considered in this analysis are consistent with current experimental conditions~\cite{Gluschke14}.

We also scrutinize the impact of quantum coherence on the driving energy $W_I$ and phononic current $Q_{\rm ph}$. 
We observe a comparative decrease for the input work $W_I$ due to the quantum coherence in Fig.~\ref{fig:WetaQph}(c). 
Similarly, the input heat current $\braket{Q_{\rm ph}}$ around $\Delta\mu=0$ in Fig.~\ref{fig:WetaQph}(d) is reduced to approximately twice the value in absence of quantum coherence.
Hence, quantum coherence enhances the thermodynamic efficiency.

\begin{figure}[htb]
\begin{center}
\centering\includegraphics[width=8.5cm]{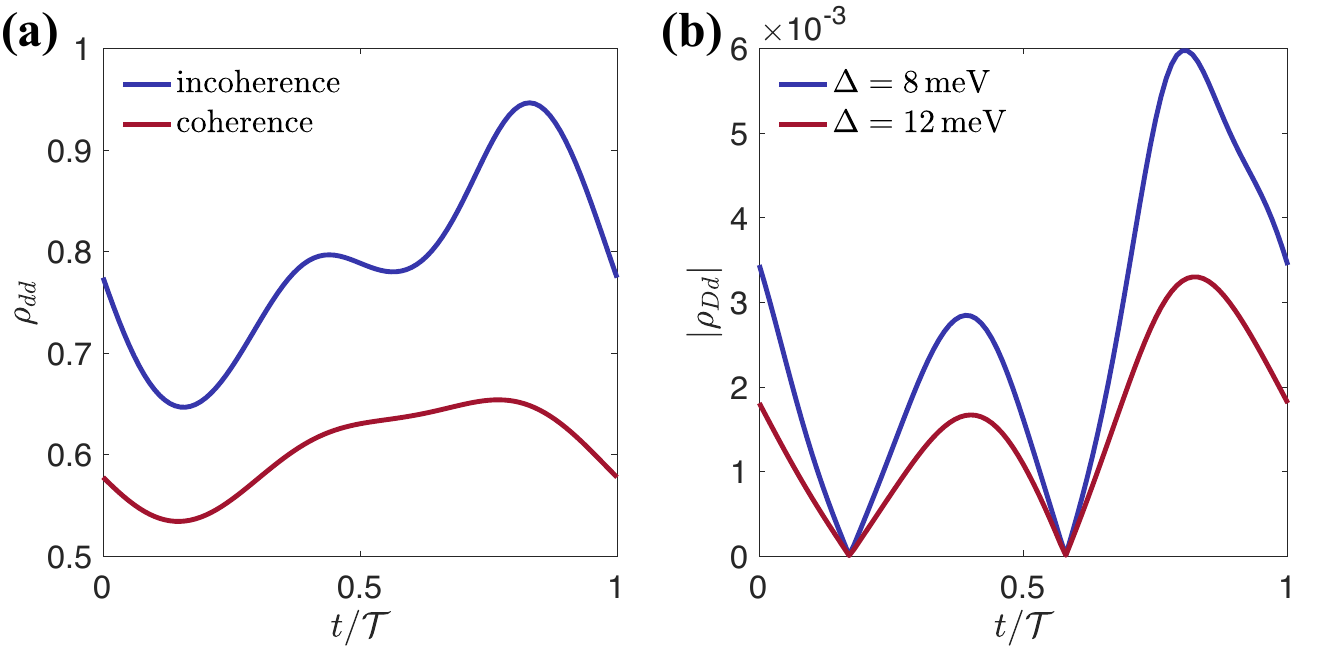}
\caption{The density matrix (a) $\rho_{dd}$ with coherence and incoherence as a function of $t/{\mathcal T}$. (b) $|\rho_{Dd}|$ as a function of $t/{\mathcal T}$ for different dot-dot tunneling strength $\Delta$. The double QDs energy modulations: $\varepsilon_l= [-1+5\sin(\Omega t)]\, {\rm meV}$, $\varepsilon_r =[1 + 5\sin(\Omega t+\pi/2)]\, {\rm meV}$, $\Omega=2\pi/{\mathcal T}$ and ${\mathcal T}=10^{-12} \,{\rm s}$. The other parameters are given by $\mu=0$, $\Gamma_l=\Gamma_r=\Gamma_{\rm ph}=4 \, {\rm meV}$, $\Delta=8 \, {\rm meV}$, $\Delta\mu = 0$, $k_BT_l=k_BT_r=10\, {\rm meV}$ and $k_BT_{\rm ph}=12 \, \rm meV$. }\label{fig:rho}
\end{center}
\end{figure}




\begin{figure}[htb]
\begin{center}
\centering\includegraphics[width=8.5cm]{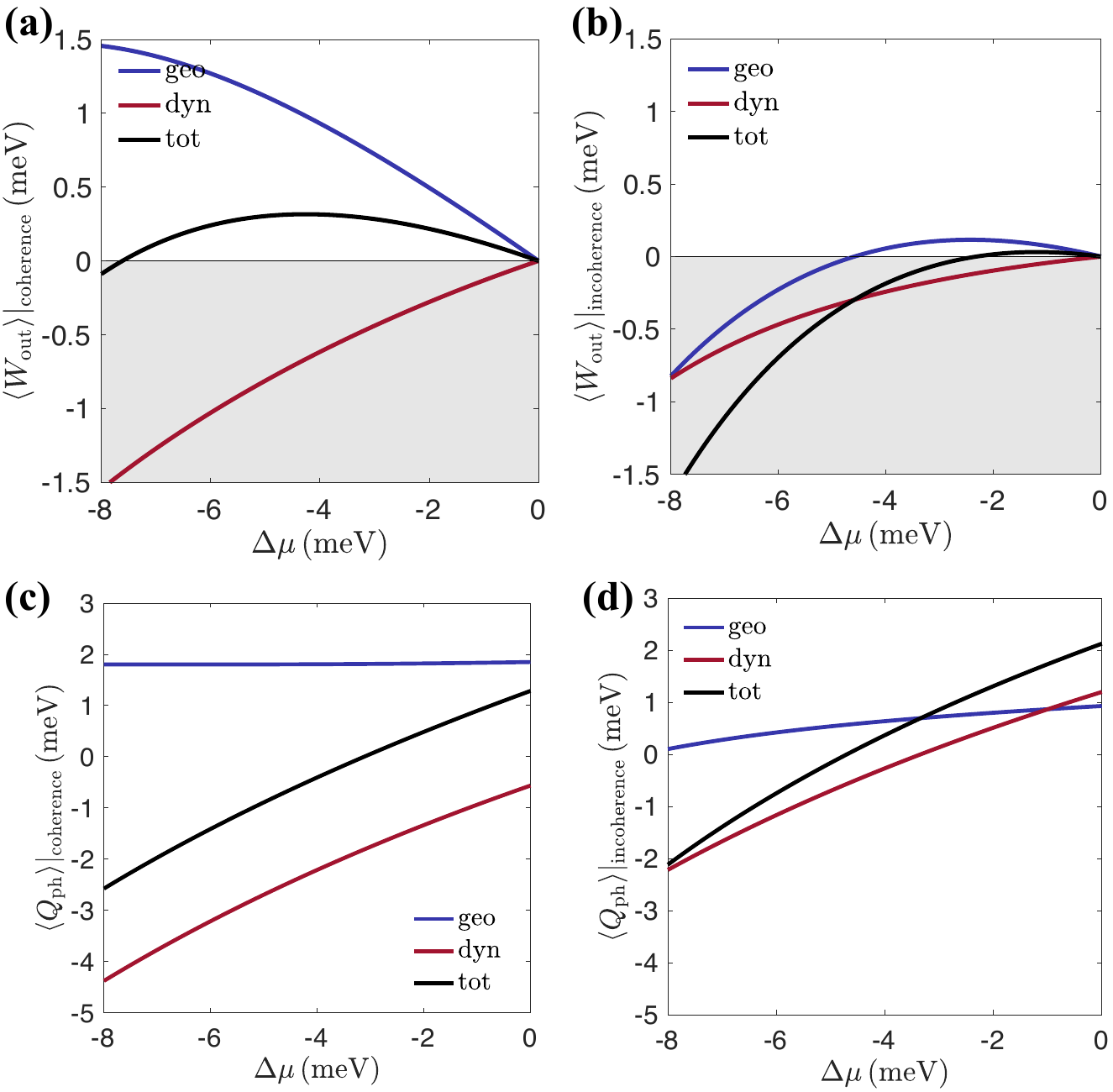}
\caption{The average output work $\braket{W_{\rm out}}$ with (a) coherence and (b) incoherence, (c) the photonic current $\braket{Q_{\rm ph}}$ with  (c) coherence and (d) incoherence as a function of $\Delta\mu$. 
The shaded  regimes in (a) and (b) denote the useless work.
The parameters are the same as in Fig.~\ref{fig:WetaQph}.} \label{fig:currentV} 
\end{center}
\end{figure}

Next, we utilize coherence measurement (i.e. $|\rho_{Dd}|$) to quantitatively estimate the quantum coherence during the adiabatic transport~\cite{PlenioPRL}. 
In Fig.~\ref{fig:rho}(a), 
it is found that the existence of $|\rho_{Dd}|$ suppresses the diagonal density matrix element $\rho_{dd}$.
And, $|\rho_{Dd}|$ itself in Fig.~\ref{fig:rho}(b) shows finite value and periodic oscillations in one driving period,
which is enhanced by the inter-dot tunneling.
Thus quantum coherence should not be naively ignored.
Considering the contribution of $|\rho_{Dd}|$ to the currents, e.g.,
the output work
$\langle W_{\rm out} \rangle=(\mu_l-\mu_r)\int_0^{\mathcal T}\frac{dt}{{\mathcal T}}\sum_{i=D,d}[-\gamma_{r,i0}\rho_{ii}(t) + \gamma_{r,0i}\rho_{00}(t)] + \frac{1}{4}\sin2\theta[{\tilde\gamma}_{r,i0}(\rho_{Dd} + \rho_{dD})]$, with the rates $\gamma_{r,i0}=\Gamma_r \lambda_{0i}[1 - f_r(\varepsilon_{i})]$, $\gamma_{r,0i}=\Gamma_r \lambda_{0i}f_r(\varepsilon_{i})$, $\tilde\gamma_{r,i0}=\Gamma_r [1 - f_r(\varepsilon_{i})]$, and $\tilde\gamma_{r,0i}=\Gamma_r f_r(\varepsilon_{i})$ ($\lambda_{0D}=\sin^2\theta$, $\lambda_{0d}=\cos^2\theta$)~\cite{BhandariPRB21,JordanPRR22},
it is known that the output work is dramatically enhanced via the quantum coherence in Fig.~\ref{fig:WetaQph}(a).
Therefore, quantum coherence indeed plays a pivotal role in the performance of quantum heat engines.


To provide a further understanding of how coherent and incoherent transport impact output work and phononic currents, we categorize these currents into two distinct components: the geometric and dynamic components~\cite{MyPRBGTUR,RenCPL23}, as elegantly illustrated in Fig.~\ref{fig:currentV}. The former component arises as a consequence of external periodic driving. In contrast, the latter one is attributed to thermodynamic biases, such as differences in chemical potentials and temperature gradients. 
It's evident that geometric work can counteract the direction of thermodynamic biases, allowing for the realization of a geometric thermoelectric pumping effect, e.g., based on the three-terminal double QDs system~\cite{MyPRBGTUR,RenCPL23}. 
The input work ${\langle}W_I{\rangle}$ at Eq.~(\ref{eq:conversation}) is completely determined by the geometric contribution, expressed as
$\langle W_I \rangle=-(\braket{ E_l}|_{\rm geo} + \langle E_r \rangle |_{\rm geo} + \langle Q_{\rm ph} \rangle |_{\rm geo})$,
based on the first law of thermodynamics,
i.e., 
$ \braket{ E_l}|_{\rm dyn} +\langle E_r \rangle |_{\rm dyn}+ \langle Q_{\rm ph} \rangle |_{\rm dyn}\equiv 0$.
The input work is apparently reduced via quantum coherence.
Hence, by comparing output work (Figs.\ref{fig:currentV}(a)-\ref{fig:currentV}(b)), heat current (Figs.\ref{fig:currentV}(c)-\ref{fig:currentV}(d)),
and input work (Fig.\ref{fig:WetaQph}(c)) both at  coherence and incoherence cases, it is interesting to note that   geometric currents experience dramatic improvement due to quantum coherence effects.
Hence, such geometric effect will strongly affect the thermodynamic performance.

\begin{figure}[htb]
\begin{center}
\centering\includegraphics[width=8.5cm]{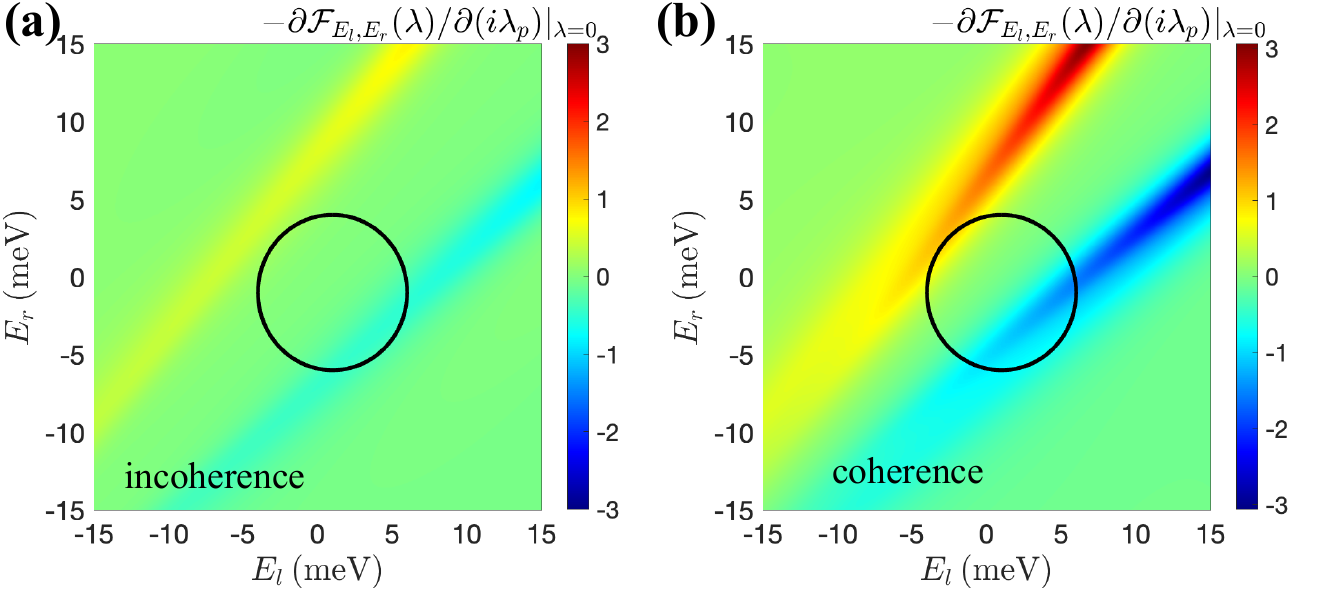}
\caption{The contour map in the parameter space of the left QD energy level $\varepsilon_l$ and the right QD energy $\varepsilon_r $: Berry curvature for the average particle current $-\partial{\mathcal F}_{\varepsilon_l \varepsilon_r }(\lambda)/\partial(i\lambda_p)|_{\lambda=0}$ (a) with incoherence and (b) coherence cases. The black circle denotes the path of two parameter QDs energy modulations: $\varepsilon_l = [-1+5\sin(\Omega t)]\, {\rm meV}$, $\varepsilon_r =[1 + 5\sin(\Omega t+\pi/2)]\, {\rm meV}$, $\Omega=2\pi/{\mathcal T}$ and ${\mathcal T}=10^{-12} \,{\rm s}$. The other parameters are given by $\mu=0$, $\Gamma_L=\Gamma_r=\Gamma_{\rm ph}=4 \, {\rm meV}$, $\Delta=8 \, {\rm meV}$, $\Delta\mu = 0$, $k_BT_l=k_BT_r=10\, {\rm meV}$ and $k_BT_{\rm ph}=12 \, \rm meV$. }\label{fig:BerryCur}
\end{center}
\end{figure}

Alternatively, in scenarios where pairs of parameters [$\varepsilon_l(t)$] and [$\varepsilon_r (t)$] are subjected to periodic driving~\cite{SanchezPRB08,LudovicoPRB18,MonselPRB22}, the gauge-invariant Berry curvature is harnessed to describe the system's geometric behavior. This Berry curvature is elegantly expressed as~\cite{bohm03,RenPRL10,RenCPL23}:
\begin{equation}
{\mathcal F}_{\varepsilon_l\varepsilon_r } = \langle \partial_{\varepsilon_l} \varphi|\partial_{\varepsilon_r } \psi\rangle - \langle \partial_{\varepsilon_r } \varphi|\partial_{\varepsilon_l} \psi\rangle,
\end{equation}
and the geometric contribution of cumulant generating function is described as 
${\mathcal G}_{\rm geo} =- \iint_{\varepsilon_l\varepsilon_r } d\varepsilon_ld\varepsilon_r  {\mathcal F}_{\varepsilon_l\varepsilon_r }$~\cite{berry84}. As shown in Fig.~\ref{fig:BerryCur}, the Berry-phase effect is able to generate geometric particle and heat currents against thermodynamic biases. 
Specifically, the Berry curvatures for the particle current, considering both incoherence [Fig.~\ref{fig:BerryCur}(a)] and coherence effects[Fig.~\ref{fig:BerryCur}(b)], promise the existence of geometric currents. 
However, quantum coherence will further enhance the performance of Berry curvature within the driving zone (rounded by black circles), 
which finally significantly strengthens geometric contribution to the thermodynamic performance of the inelastic heat engine. 

\section{conclusion}~\label{sec:con}

In summary, we have demonstrated that the quantum coherence  can enhance the thermodynamic performance of periodically-driven quantum heat engines. Employing quantum master equations with the full counting statistics method, which conserves quantum coherence, we derived the dynamics and geometric current of coherent quantum heat engine, along with the output work and thermodynamic efficiency. Our results reveal that inelastic quantum heat engine exhibits significantly higher performance compared to their elastic counterpart. 
The nonzero quantum coherence, characterized as the off-diagonal density matrix elements in eigenbasis, is explicitly exhibited.
Various modulations of system parameters, e.g., geometric modulation phase and tunneling strength, on thermodynamic performance are exhbited in Appendix \ref{sec:appC}. 

Furthermore, we delved into exploring the impact of quantum coherence on the thermodynamic performance of the periodically driven inelastic heat engine. 
Through external modulations with dual parameters and considering the geometric phase, we illustrated that the quantum coherence effect can enhance heat absorption and inelastic currents, consequently improving output work and thermodynamic efficiency. Analyzing the Berry curvature, we further unveiled the mechanism of quantum coherence on geometric currents. 
It is shown that the curvature is dramatically strengthened via quantum coherence.
Therefore, our findings provide physical insights for optimizing periodically-driven inelastic heat engines with quantum coherence resource.

\section{Acknowledgement}  
This work was supported by the funding for the National Natural Science Foundation of China under Grants No. 12125504, No. 12074281, No. 11935010 and No. 12305050, Jiangsu Key Disciplines of the Fourteenth Five-Year Plan (Grant No. 2021135), the Natural Science Foundation of Jiangsu Higher Education Institutions of China (Grant No. 23KJB140017), and the Opening Project of Shanghai Key Laboratory of Special Artificial Microstructure Materials and Technology.

\appendix 

\begin{widetext}

\section{Full Counting Statistics for Particle and Energy Currents} \label{sec:FCS}

We briefly introduce the full counting statistics to count the particle and energy flows in the inelastic heat engine.
Here, we count the particle and energy flows into the $r$-th electronic reservoir and the heat current into the $ph$-th reservoir.
Using the two-time measurement protocol~\cite{fluctuationRMP, CampisiRMP},
one is able to define the characteristic function  as
\begin{equation}
\mathcal{Z}(\mathbf{\Lambda})=\langle e^{i\lambda_p\hat{N}_p(0)+i\lambda_E\hat{H}_E(0)+i\lambda_{\rm ph}\hat{H}_{\rm ph}(0)}e^{-i\lambda_p\hat{N}_p(t)-i\lambda_E\hat{H}_E(t)-i\lambda_{\rm ph}\hat{H}_{\rm ph}(t)}\rangle,
\label{eq:parfun}
\end{equation}
where $\mathbf{\Lambda}=\{\lambda_p,\lambda_E,\lambda_{\rm ph}\}$,
$\lambda_{p,E}$ counting parameters for particle
and energy flows into the $r$-th reservoir, and 
$\lambda_{\rm ph}$ heat currents into phonon reservoir, respectively.  
$\hat{N}_p$ represents the number operator for the total particles in the $r$-th reservoir, $\hat{H}_E$ is the Hamiltonian operator for the $r$-th reservoir, and $\hat{H}_{\rm ph}$ is the Hamiltonian operator for the phonon reservoir. 
The time evolution follows the Heisenberg representation, and $\langle .\rangle$ denotes an average with respect to the total initial density operator. This density operator is considered as a factorized form with respect to the central QDs system ($S$) and the ($l$, $r$, and $\rm ph$) reservoirs. It is expressed as 
\begin{eqnarray}
\rho_T(0) = \rho_S(0) \otimes \rho_l(0) \otimes \rho_r(0) \otimes \rho_{\rm ph}(0),
\end{eqnarray}
where the equilibrium distribution of $\rho_{\alpha}$ of the $\alpha$-th electronic reservoir is given by $\rho_{\alpha} = \exp\left[-\beta_{\alpha}(\hat H_{\alpha} - \mu_\alpha\hat N_\alpha)\right] / Z_\alpha$, with $Z_\alpha = \text{Tr}\left\{\exp\left[-\beta_{\alpha}(\hat H_{\alpha} - \mu_\alpha\hat N_\alpha)\right]\right\}$ representing the partition function,
and equilibrium density operator of the phonon reservoir is expressed as
$\rho_{\rm ph}(0)=\exp\left[-\beta_{\rm ph}\hat H_{\rm ph} \right] / Z_{\rm ph}$, with 
$Z_{\rm ph} = \text{Tr}\left\{\exp\left[-\beta_{\rm ph}\hat H_{\rm ph} \right]\right\}$.
Actually, Eq.~\eqref{eq:parfun} can be reorganized as  
\begin{eqnarray}
\mathcal{Z}_t(\mathbf{\Lambda})={\rm Tr}[\rho^T_{\mathbf{\Lambda}}(t)],
\end{eqnarray}
where the modified density operator is specified as
\begin{eqnarray}
\rho^T_{\mathbf{\Lambda}}(t)=U_{-\mathbf{\Lambda}/2}(t)\rho_T(0)U^{\dagger}_{\mathbf{\Lambda}/2}(t).
\label{eq:A4}
\end{eqnarray}
The  propagating operator embedded with counting parameters is given by
$U_{-\mathbf{\Lambda}/2}(t) 
= e^{-i\frac{\lambda_{\rm ph}}{2}H_{\rm ph}-i\frac{\lambda_E}{2}H_r-i\frac{\lambda_p}{2}N_r}U(t)e^{i\frac{\lambda_{\rm ph}}{2}H_{\rm ph}+i\frac{\lambda_E}{2}H_r+i\frac{\lambda_p}{2}N_r}$,
which can be reexpressed as
\begin{eqnarray}
 U_{-\mathbf{\Lambda}/2}(t)=e^{-iH_{-\mathbf{\Lambda}/2}(t)},     
\end{eqnarray}
with the counting-field-dependent total Hamiltonian   
\begin{eqnarray}
H_{-\mathbf{\Lambda}/2} 
= e^{-i\frac{\lambda_{\rm ph}}{2}H_{\rm ph}-i\frac{\lambda_E}{2}H_r-i\frac{\lambda_p}{2}N_r}H_Te^{i\frac{\lambda_{\rm ph}}{2}H_{\rm ph}+i\frac{\lambda_E}{2}H_r+i\frac{\lambda_p}{2}N_r},
\end{eqnarray}
and $H_T$ is the total Hamiltonian of the inelastic heat engine.
Hence, the $t$-time cumulant generting function is given by
$G_t(\mathbf{\Lambda})
={\partial}\ln\mathcal{Z}_t(\mathbf{\Lambda})/{\partial}t$.
Then the current can be obtained as
\begin{eqnarray}
J_\mu(t)=\frac{{\partial}G_t(\mathbf{\Lambda})}{{\partial}(i\lambda_\mu)}\Big{|}_{\mathbf{\Lambda}=0}.
\end{eqnarray}

\newpage

\section{The detailed expression of the evolution ${\mathbf H}(\Lambda)$ for counting the right reservoir}~\label{countright}

The following is the detailed expression of the evolution ${\mathbf H}(\bf \Lambda)$ for counting the right reservoir:

\begin{equation}
\begin{aligned}
H_{11}(\bf \Lambda)&=-\Gamma_l \cos^2\theta f_l (E_D) -\Gamma_l \sin^2\theta f_l (E_d) - \Gamma_r \sin^2\theta f_r (E_D) - \Gamma_r \cos^2\theta  f_r (E_d),\\ 
H_{12}(\bf \Lambda)& = \Gamma_l  \cos^2\theta [1 - f_l(E_D)] + \Gamma_r  \sin^2\theta [1-f_r(E_D)]  e^{-i(\lambda_p + \lambda_E E_D)},\\
H_{13}(\bf \Lambda)& = \Gamma_l \sin^2\theta [1-f_l(E_d)] + \Gamma_r \cos^2\theta [1 - f_r(E_d)] e^{-i(\lambda_p + \lambda_E E_d)} ,\\
H_{14}(\bf \Lambda)& = H_{15}(\mathbf{\Lambda}) = \frac{1}{2}\Gamma_l \sin\theta \cos\theta [[1-f_l(E_D)] + [1- f_l(E_d)] ] \\
&\quad\quad\quad- \frac{1}{2}\Gamma_r \sin\theta \cos\theta [[1-f_r(E_D)] e^{-i(\lambda_p + \lambda_E E_D)} + [1 - f_r(E_d)] e^{-i(\lambda_p + \lambda_E E_d)}],\\ 
H_{21}(\bf \Lambda)& = \Gamma_l \cos^2\theta f_l(E_D) + \Gamma_r \sin^2\theta f_r(E_D) e^{i(\lambda_p + \lambda_E E_D)},\\
H_{22}(\bf \Lambda)& = -\Gamma_l \cos^2\theta [1-f_l(E_D)] - \Gamma_r \sin^2\theta [1-f_r(E_D)]-\Gamma_{\rm ph} \cos^2 2\theta \, [1 + n(\omega_0)], \\
H_{23}(\bf \Lambda)& = \Gamma_{\rm ph} \cos^2 2\theta e^{i\lambda_{\rm ph}\omega_0} \, n(\omega_0),\\ 
H_{24}(\bf \Lambda)& =H_{25}(\mathbf{\Lambda})= \frac{1}{2} \Gamma_l \sin\theta \cos\theta [1 - f_l(E_d)] - \frac{1}{2}\Gamma_r \sin\theta \cos\theta [1 - f_r(E_d)]+ \frac{1}{2}\Gamma_{\rm ph} \sin2\theta \cos2\theta (e^{i\lambda_{\rm ph}\omega_0} -1) \, [1 + n(\omega_0)],\\    
H_{31}(\bf \Lambda)& =  \Gamma_l  \sin^2\theta f_l(E_d) + \Gamma_r \cos^2\theta f_r(E_d) e^{i(\lambda_p + \lambda_E E_d)},\\
H_{32}(\bf \Lambda)& =  \Gamma_{\rm ph} \cos^2 2\theta \, e^{-i\lambda_{\rm ph}\omega_0} \, [1 + n(\omega_0)],\\   
H_{33}(\bf \Lambda)& = -\Gamma_l \sin^2\theta [1-f_l(E_d)] - \Gamma_r \cos^2\theta [1-f_r(E_d)] - \Gamma_{\rm ph} \cos^2 2\theta \, n(\omega_0),\\
H_{34}(\bf \Lambda)& =H_{35}= \frac{1}{2}\Gamma_l \sin\theta \cos\theta [1 - f_l(E_D)] - \frac{1}{2}\Gamma_r \sin\theta \cos\theta [1 - f_r(E_D)]-\frac{1}{2}\Gamma_{\rm ph} \sin2\theta \cos2\theta (e^{-i\lambda_{\rm ph}\omega_0} - 1)\,[1 + n(\omega_0)], \\   
H_{41}(\bf \Lambda)& =H_{51}=  -\frac{1}{2}\Gamma_l \sin\theta \cos\theta [f_l(E_D) + f_l(E_d) ] + \frac{1}{2}\Gamma_r \sin\theta \cos\theta [f_r(E_D)e^{i(\lambda_p + \lambda_E E_D)}  + f_r(E_d) e^{i(\lambda_p + \lambda_E E_d)}], \\   
H_{42}(\bf \Lambda)& = H_{52}(\mathbf{\Lambda})=\frac{1}{2}\Gamma_l \sin\theta \cos\theta [1 - f_l(E_D)] - \frac{1}{2}\Gamma_r \sin\theta \cos\theta [1 - f_r(E_D)] + \frac{1}{2}\Gamma_{\rm ph} \sin2\theta \cos2\theta (1 + e^{-i\lambda_{\rm ph}\omega_0}) \, [1+n(\omega_0)],\\
H_{43}(\bf \Lambda)& =H_{53}(\mathbf{\Lambda})= \frac{1}{2}\Gamma_l\sin\theta\cos\theta [1-f_l(E_d)] - \frac{1}{2}\Gamma_r \sin\theta\cos\theta [1-f_r(E_d)] -\frac{1}{2}\Gamma_{\rm ph} \sin2\theta \cos2\theta  (e^{i\lambda_{\rm ph}\omega_0} + 1)\, n(\omega_0),\\
H_{44}(\bf \Lambda)& =H_{55}(\mathbf{\Lambda})= -\frac{1}{2}\Gamma_l \sin^2\theta [1-f_l(E_D)] -\frac{1}{2}\Gamma_l \cos^2\theta [1-f_l(E_d)] - \frac{1}{2}\Gamma_r \cos^2\theta [1-f_r(E_D)] - \frac{1}{2}\Gamma_r \sin^2\theta [1-f_r(E_d)] \\ 
&\quad\quad\quad \quad- \frac{1}{2}\Gamma_{\rm ph}\cos^22\theta [1 + 2n(\omega_0)],\\
H_{45}(\mathbf{\Lambda})& =H_{54}(\mathbf{\Lambda})= \frac{1}{2}\Gamma_{\rm ph}\cos^22\theta e^{-i\lambda_{\rm ph}\omega_0} \, [1 + n(\omega_0)] + \frac{1}{2}\Gamma_{\rm ph}\cos^22\theta  e^{i\lambda_{\rm ph}\omega_0} \, n(\omega_0).  
\end{aligned}
\end{equation}
The counting parameters set denotes $\mathbf{\Lambda}=\{\lambda_p,\lambda_E,\lambda_{\rm ph}\}$, $\Gamma_{i}=2\pi\sum_k |\gamma_{i,k}|^2 \delta(E-\varepsilon_{i,k})$ is the dot-electronic reservoir hybridization energy, and $\Gamma_{\rm ph}=2\pi \sum_q {\lambda^2_q} \delta(\omega-\omega_q)$ is the coupling energy of the phonon bath. $f_i(\varepsilon_i)=\{\exp[(\varepsilon_i-\mu_i)/k_BT_i] +1\}^{-1}$ is the Fermi-Dirac distribution for the electronic reservoir with chemical potential $\mu_i$ and the temperature $T_i$, and $n(\omega_0)=[\exp(\omega_0/k_BT_{\rm ph})-1]^{-1}$ is the Bose-Einstein distribution function in the phonon reservoir with the temperature $T_p$
and energy gap $\omega_0=E_D-E_d$.

\begin{figure}[htb]
\begin{center}
\centering\includegraphics[width=17.0cm]{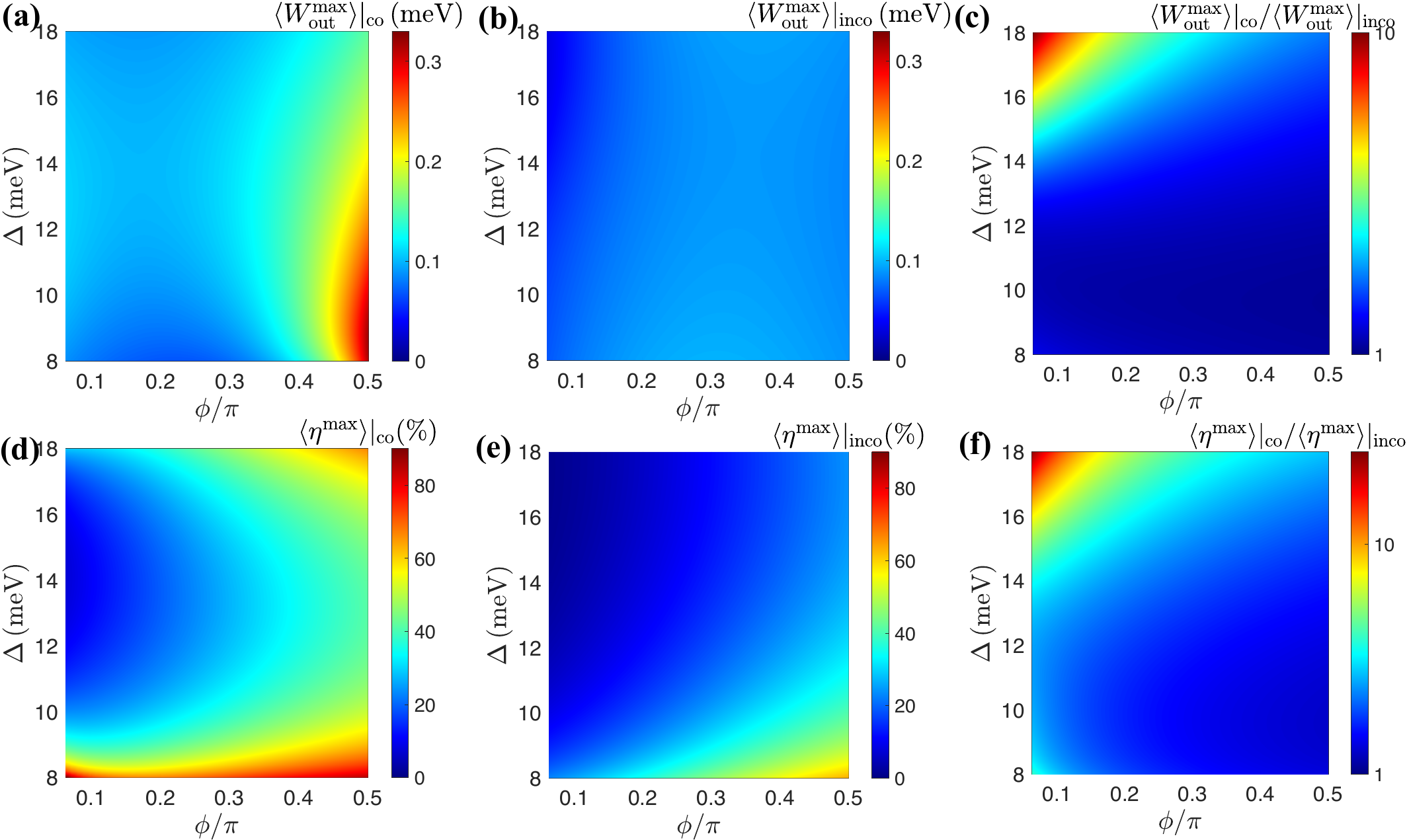}
\caption{(a) The coherent maximum output work $\braket {W^{\max}}|_{\rm co}$, (b) the  maximum output work $\braket {W^{\max}}|_{\rm inco}$, (c) the ratio of the coherent and incoherent maximum work $\braket {W^{\max}}|_{\rm co}/\braket {W^{\max}}|_{\rm inco}$, (d) the coherent maximum efficiency $\braket {\eta}^{\max}|_{\rm co}$, (e) the incoherent maximum efficiency $\braket {\eta}^{\max}|_{\rm co}$, (f) the ratio of the coherent and incoherent maximum efficiency ratio $\braket {\eta^{\max}}|_{\rm co}/\braket {\eta^{\max}}|_{\rm inco}$ vs modulation phase $\phi$ and tunneling strength $\Delta$. The parameters are given by $\mu=0$, $\Gamma_l=\Gamma_R=\Gamma_{\rm ph}=4 \, {\rm meV}$, $k_BT_l=k_BT_r=10\, {\rm meV}$ and $k_BT_{\rm ph}=12 \, \rm meV$. The energy modulations are exemplified as $\varepsilon_l = [-1+5\sin(\Omega t)]\, {\rm meV}$, $\varepsilon_r =[1 + 5\sin(\Omega t+\pi/2)]\, {\rm meV}$, $\Omega=2\pi/{\mathcal T}$ and ${\mathcal T}=10^{-12} \,{\rm s}$, $\mu=0$.   }   
\label{fig:MaxWeta}
\end{center}
\end{figure}

\section{Performance of the inelastic coherently engines with different modulation phase, coherent tunneling strength}\label{sec:appC}

 Here, we aim to investigate the impact of various parameters, such as the tunneling strength between the double quantum dots and the modulation phase, on  efficiency and output work, without making specific parameter selections. To begin, we explore how coherent transport behaves with respect to these parameters. In Fig.~\ref{fig:MaxWeta}, we present the average output work and efficiency of the  heat engine as functions of the tunneling strength ($\Delta$) and the phase ($\phi$). For each configuration, we optimize performance by adjusting the chemical potential difference (i.e., the chemical potential difference $\Delta\mu$ at maximum efficiency and work). Fig.~\ref{fig:MaxWeta}(a) reveals that coherent transport has a pronounced impact on the maximum output work. For instance, with $\phi=\pi/2$ and $\Delta = 8\, \rm meV$, the maximum work increases from $0.1\, \rm meV$ to $0.4\, \rm meV$, due to quantum coherence. We also calculate the enhancement ratio $\braket {W^{\max}}|_{\rm co}/\braket {W^{\max}}|_{\rm inco}$ for different tunneling strengths and modulation phases, as shown in Fig.~\ref{fig:MaxWeta}(c). This enhancement ratio can be as high as $10$, indicating a substantial improvement in work with increasing tunneling strength and decreasing phase.

Moreover, as shown in Fig.~\ref{fig:MaxWeta}, although the modulation phase $\phi$ can increase the magnitude of the geometric current, it doesn't consistently improve the overall performance of heat engines. This discrepancy can be attributed to the potential inconsistency between the direction of geometric and dynamical currents, which can lead to a reduction of the total magnitude of nonequilibrium currents in the system. Meanwhile, the efficiency is optimized when the phase is either small or large, as illustrated in Figs.~\ref{fig:MaxWeta}(c) and \ref{fig:MaxWeta}(d). Additionally, the tunneling strength between the quantum dots plays an important role in enhancing quantum coherence. As shown in Figs.~\ref{fig:MaxWeta}(a) and \ref{fig:MaxWeta}(b).  
These results emphasize the crucial role of quantum coherence in inelastic heat engines.

\end{widetext}

\bibliography{RefDQD}

\end{document}